\documentclass[apm,reprint,amsmath,amssymb]{revtex4-1}

\usepackage{graphicx}
\usepackage{bm}

\begin{document}

\title{Evidence for proximity of YFe$_2$Si$_2$ to a magnetic quantum critical
point}

\author{David J. Singh}

\affiliation{
Department of Physics and Astronomy, University of Missouri,
Columbia, MO 65211-7010, USA}

\email{singhdj@missouri.edu}

\date{\today}

\begin{abstract}
Calculations of the electronic and magnetic properties of the
non-magnetic metallic compound YFe$_2$Si$_2$ are reported.
These show that at the
density functional level a magnetic state involving ordering along the
$c$-axis. The electronic structure is three dimensional, and is similar
to that of the unconventional superconductor YFe$_2$Ge$_2$ and as well
as that of the
high pressure collapsed tetragonal phase of KFe$_2$As$_2$, which is
also a superconductor. Based on the results in relation to experiment,
we infer that properties of YFe$_2$Si$_2$ are strongly influenced by
a nearby antiferromagnetic quantum critical point.
\end{abstract}


\maketitle

\section{Introduction}

There is renewed interest in metals with unusual magnetic behavior.
This is due in part to the
unusual magnetic properties of the Fe-based superconductors,
\cite{johnston,stewart,lumsden,mazin-mag,bondino}
and spin-fluctuation pairing models for these and other unconventional
superconductors.
\cite{scalapino,mathur,johannes,moriya,mazin-spm,kuroki-spm,scalapino2}
Within such models superconductivity depends both on the details of the
spin-fluctuations, particularly their strength and momentum dependence,
and the electronic structure, i.e. the Fermi surface and the coupling
of states on it with the spin fluctuations.
In Fe-based superconductors, the electronic structure involves several
$d$ orbitals, hybridized with ligand p-states, and formed
by both hopping through ligand orbitals and direct Fe-Fe hopping.
In most cases,
this leads to disconnected hole and electron sheets of Fermi surface
connected by antiferromagnetic fluctuations associated with a stripe 
magnetic order. \cite{mazin-spm,kuroki-spm,singh-du}
However, there are heavily electron doped materials, such as K$_x$Fe$_2$Se$_2$,
that apparently only have electron Fermi surfaces, but are still
high temperature superconductors.
\cite{guo,fang,zhang-tl,qian,ye}
This both presents a challenge for
theory and suggests exploration for other compounds that might show
different forms of Fe-based superconductivity.

While Fe-based superconductivity has generally been restricted to
pnictides and chalcogenides,
Zou, Chen and co-workers, \cite{zou,chen}
have recently reported superconductivity in the germanide,
YFe$_2$Ge$_2$ with $T_c$=1.8 K.
There is strong evidence both from experiment \cite{zou,chen} and theory
\cite{subedi-yfe2ge2,singh-yfe2ge2} 
that the superconductivity is unconventional,
but the symmetry of the superconducting state has not been established.
In any case, this first finding of Fe-based superconductivity in a compound
with a group IV ligand suggests exploration of chemically related compounds.
Here we report investigation of the silicide, YFe$_2$Si$_2$.

YFe$_2$Si$_2$ is an Fe-based compound that occurs in the ThCr$_2$Si$_2$
structure, \cite{rossi}
is metallic and does not show ordered magnetism in experiment.
However, while there is no magnetic ordering, there a significant variation
in the reported magnetic properties perhaps due to sample differences.
\cite{umarji,noakes,bara,dommann,ijjaali,sankar,felner-ssc,felner-2,felner-3}

\section{Methods and Structure}

The present density functional theory (DFT) calculations were done using
the generalized gradient approximation of Perdew, Burke and Ernzerhof (PBE)
\cite{pbe} and the linearized augmented planewave (LAPW)
method \cite{singh-book} as implemented in the WIEN2k code.
\cite{wien2k}
LAPW sphere radii of 2.4 bohr, 2.4 bohr and 1.85 bohr were used for
Y, Fe and Si respectively. Well converged basis sets consisting of
local orbitals for the upper core states of Y and Fe and LAPW functions
up to a cutoff determined by $R_{Si}k_{max}$=7, corresponding to an effective
$R_{Fe}k_{max}\simeq$9 for the metal atoms, were used.
The calculations were based on the experimental lattice parameters,
$a$=3.92 \AA, $c$= 9.92 \AA. \cite{rossi}
The internal coordinate, corresponding to the Si height above the Fe plane,
was determined by energy minimization as discussed below.

As mentioned, the internal coordinate of Si in the unit cell was
determined by energy minimization. Within our density functional
calculations we find a magnetic ground state in contrast to 
experiment. We did the relaxation both for a non-spin-polarized case
and for ferromagnetic order.
With ferromagnetic order we obtain a Si position, $z_{\rm Si}$=0.3710,
as compared to 0.3673 without spin polarization.
Thus including ferromagnetic
order increases the Fe-Si distance from 2.280 \AA{} to 2.298 \AA. Unless,
noted otherwise the results presented below are based on the
ferromagnetic Si position.
Magnetism is predicted in the DFT calculations independent of this 
choice of Si position.
We note that there is a similar but larger effect of magnetism
on the structure, particularly the ligand height, in the Fe-based
superconductors. This includes compositions that are not magnetically ordered
in experiment. In the Fe-based superconductors
the best agreement with the experimental
structure is obtained from magnetic calculations. \cite{mazin-mag}

\section{Results}

\begin{figure}
\includegraphics[width=\columnwidth]{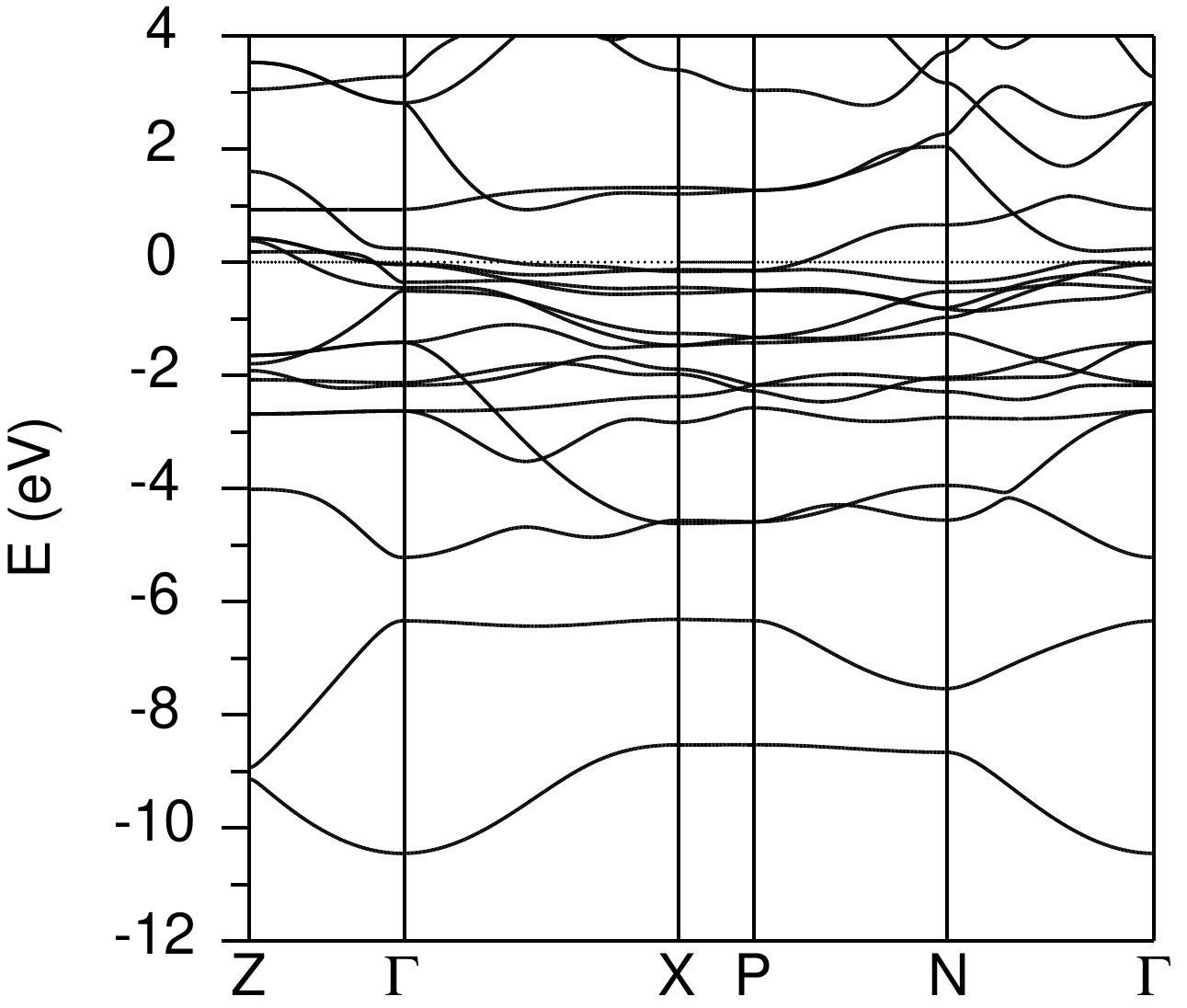}
\caption{Calculated non-spin-polarized band structure of YFe$_2$Si$_2$.}
\label{bands}
\end{figure}

\begin{figure}
\includegraphics[width=\columnwidth]{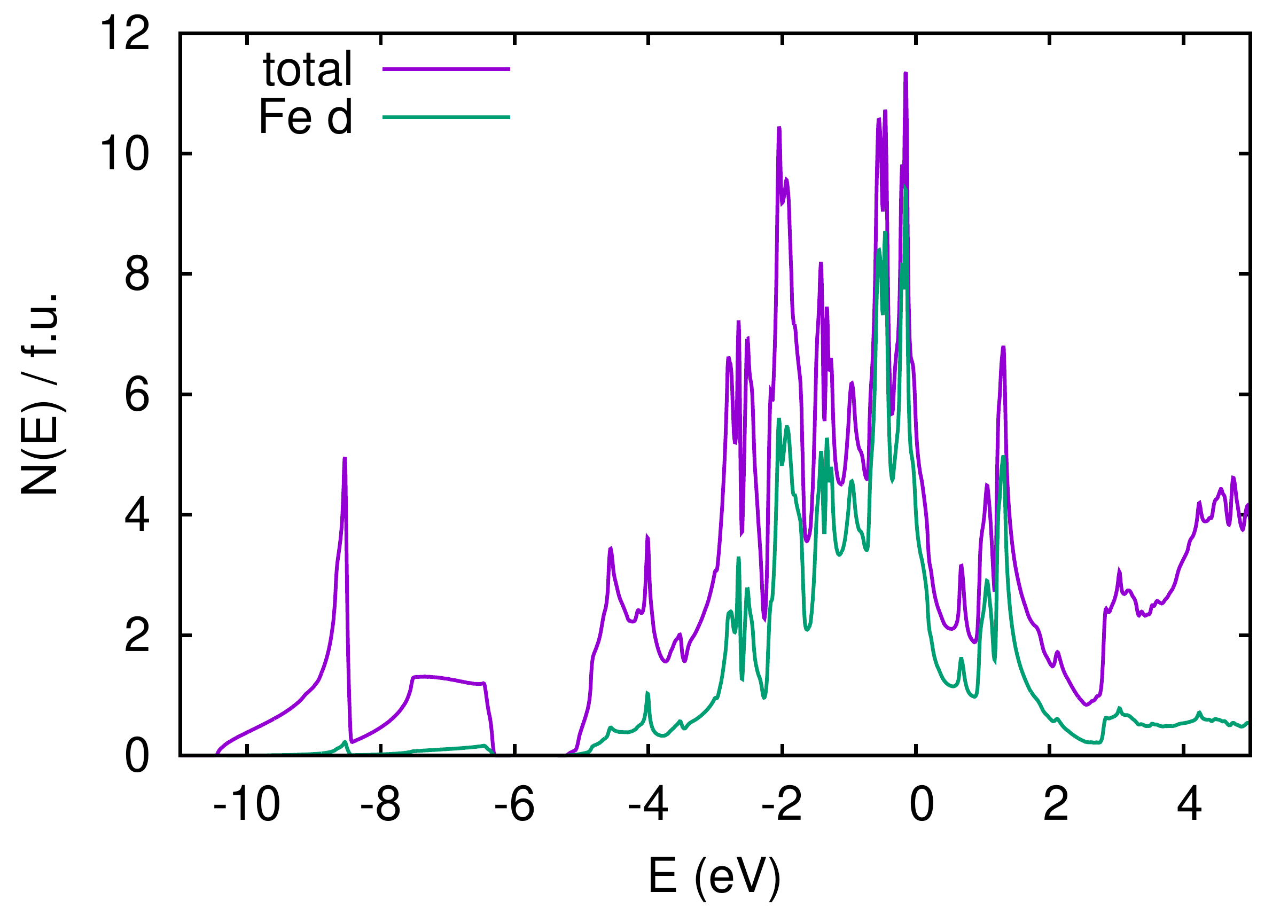}
\caption{Calculated electronic density of states and Fe d projection
on a per formula unit basis (note that there are two Fe atoms per
formula unit).}
\label{dos}
\end{figure}

\begin{figure}
\includegraphics[width=\columnwidth]{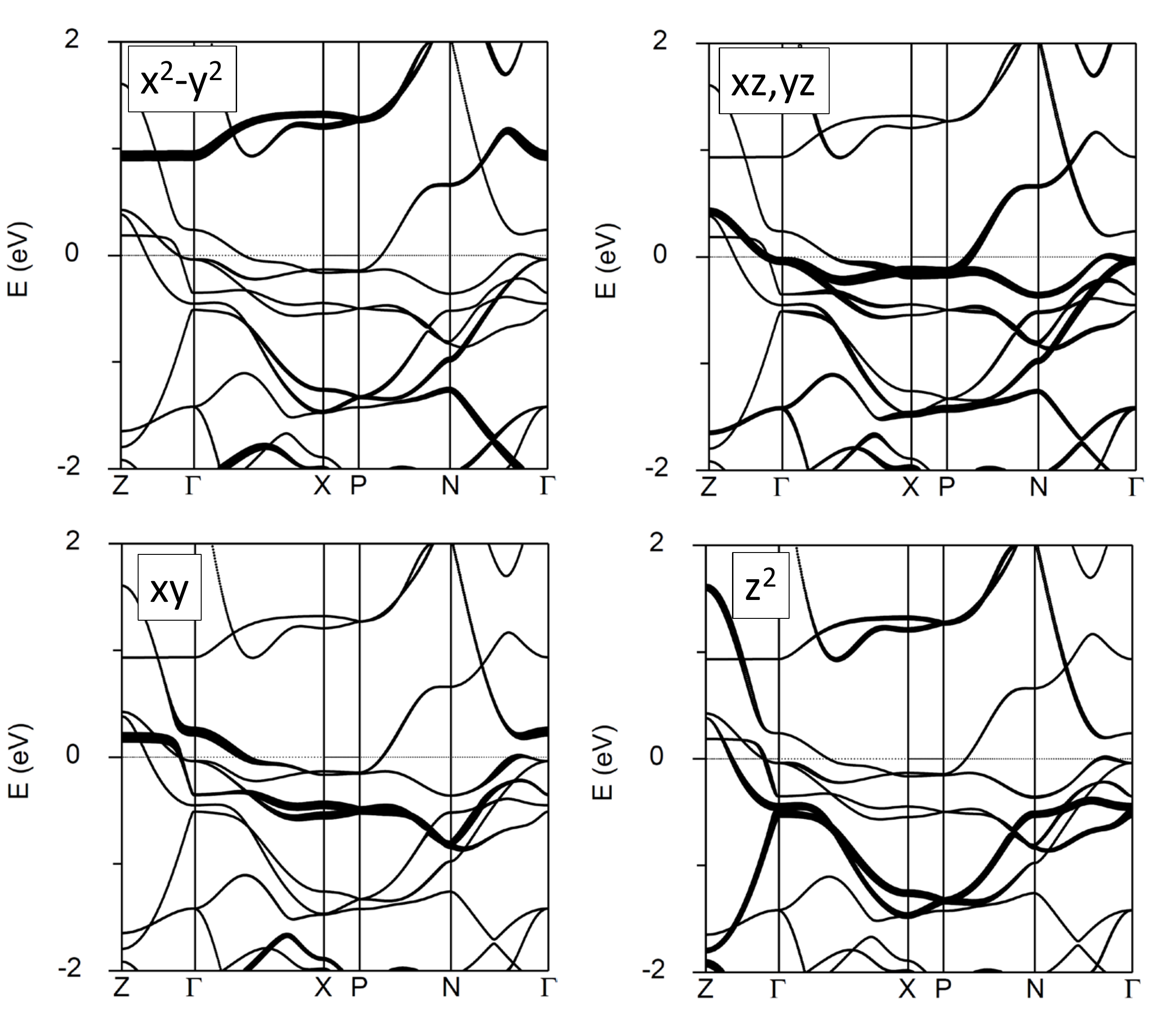}
\caption{Bands near the Fermi energy, emphasizing different Fe d
orbital characters by fat bands. The coordinate systems for the
orbital character is the square lattice defined by the Fe plane,
which is rotated 45$^\circ$ from the tetragonal $a$ and $b$ lattice
directions for the two Fe atom unit cell.}
\label{character}
\end{figure}

We start by discussing the electronic structure as obtained without
spin-polarization. The calculated band structure is shown in
Fig. \ref{bands}, with the corresponding electronic density of states
and projection of Fe $d$ character in Fig. \ref{dos}.
The lowest two bands in Fig. \ref{bands}, ($\sim$-11 eV -- -6 eV),
are from the Si $s$ orbitals, while the higher valence bands shown
are derived from Si $p$ and Fe $d$ orbitals.
The bands from -3 eV to 2 eV (relative to the Fermi energy, $E_F$) have
predominant Fe character, hybridized with Si. Y occurs as Y$^{3+}$,
with the Y $4d$ states located entirely above $E_F$ as may be expected.
The orbital characters of the Fe $d$ bands around $E_F$ are shown in
the fat band plots of Fig. \ref{character}.

\begin{figure}
\includegraphics[width=\columnwidth]{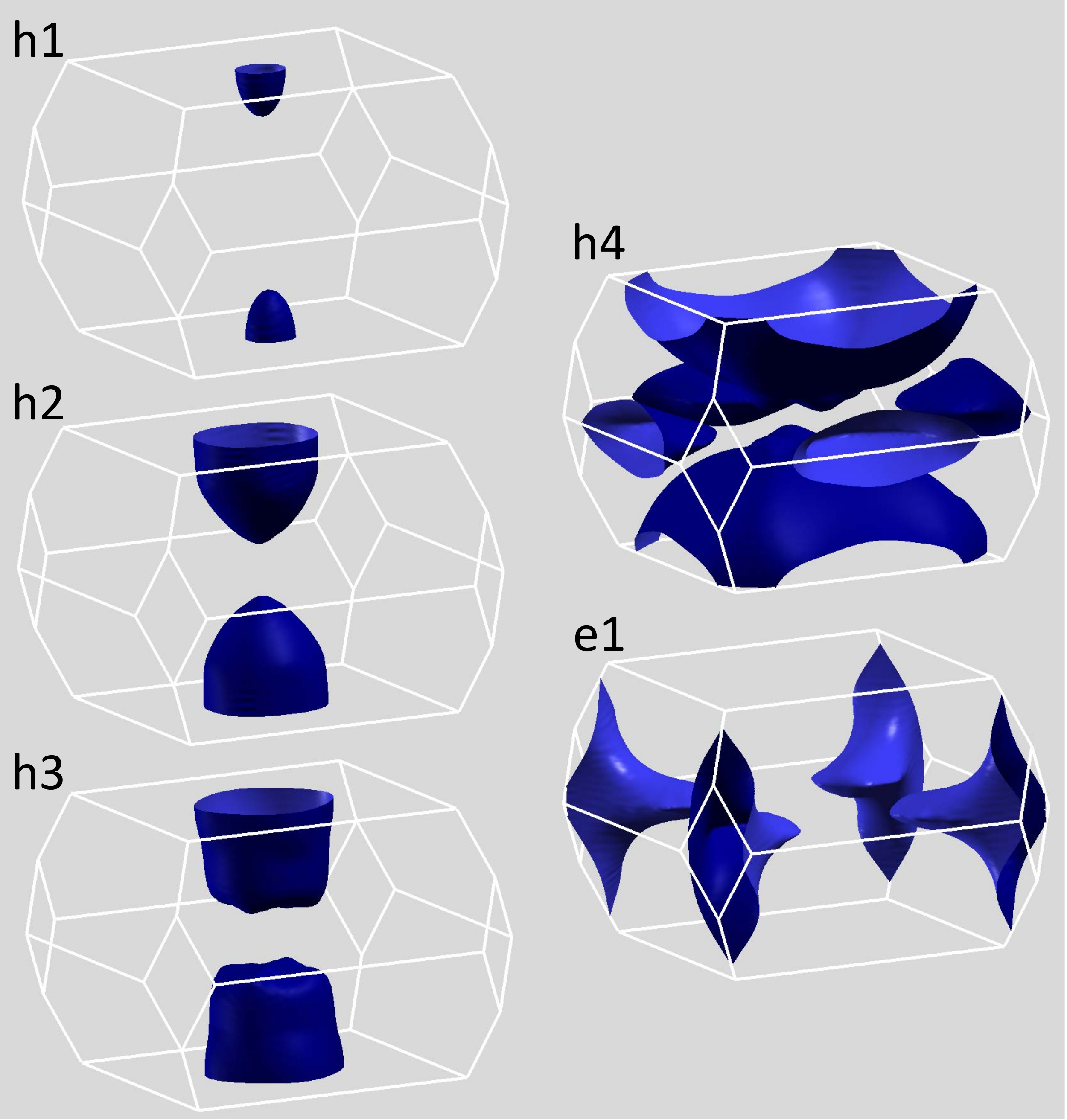}
\caption{Fermi surfaces of YFe$_2$Si$_2$, showing four hole sheets
(h1,h2,h3,h4) and one electron sheet (e1).}
\label{fermi}
\end{figure}

As seen, there are five bands crossing the Fermi level.
The resulting five Fermi surfaces are depicted in Fig. \ref{fermi}.
These involve multiple $d$ orbitals.
The density of states at the Fermi level is $N(E_F)$=5.47 eV$^{-1}$
per formula unit (two Fe atoms). 
The corresponding bare specific heat coefficient is
$\gamma_{bare}$=12.9 mJ/mol K$^2$.
It will be interesting to compare this with experiment to determine
the specific heat renormalization, which is 
$\gamma/\gamma_{bare}\sim$10 in the
germanide superconductor, YFe$_2$Ge$_2$. \cite{chen}
Since $N(E_F)$ comes from $d$ bands, the Stoner criterion for itinerant
magnetism
\cite{stoner,janak}
is clearly exceeded, and so the non-spin-polarized
state is unstable against magnetism at the DFT level.

The calculated plasma frequencies are
$\Omega_{p,xx}$=$\Omega_{p,yy}$=3.33 eV and $\Omega_{p,zz}$=4.59 eV.
Based on this, YFe$_2$Si$_2$ is an anisotropic
three dimensional metal.
This means that from an electronic point of view the material is well
connected along the $c$-axis direction. This reflects
Si-Si bonding, analogous to the Ge-Ge bonding that has been
discussed in YFe$_2$Ge$_2$, \cite{subedi-yfe2ge2,singh-yfe2ge2,chen}
and As-As bonding the collapsed tetragonal phase of KFe$_2$As$_2$.
\cite{chen}
Assuming that the scattering is relatively
k-independent, this would imply a high conductivity direction along $c$,
with an anisotropy $\sigma_c$/$\sigma_a$$\sim$1.9.
However, as discussed
below, the calculations suggest strong spin fluctuations coupled to parts of
the Fermi surface, which may produce different scattering on different sheets
and a temperature dependent anisotropy, reflecting the evolution of the
spin fluctuations with temperature.

As mentioned, there are five bands crossing $E_F$. 
These lead to five sheets of Fermi surface, consisting of four closed
hole sections, and an open corrugated
cylindrical electron section along the zone corners,
as depicted in Fig. \ref{fermi}
The electron count in YFe$_2$Si$_2$ is odd, so the electron and
hole sections are not compensated, and the hole sections
are dominant. 
The hole sheets consist of four closed sections, h1, h2, h3, and h4,
centered at the $Z$ point, and containing 
0.008, 0.107, 0.154 and 0.893 holes per formula unit, respectively.
The electron cylinder at the zone corner (e1)
contains 0.161 electrons per formula unit.
The orbital character of the small h1 section is $z^2$ (here
we use the coordinate system of the Fe-square plane, which is rotated
45$^\circ$ with respect to the $a$ and $b$ axes of the two Fe atom
unit cell), while the other three hole sections have mixed 
character involving all the $d$ orbitals except $x^2-y^2$.
The electron cylinder has predominantly $xz$,$yz$ and $xy$ character.

\begin{table}
\caption{Calculated energies and total density of states, $N(E_F)$ for
different magnetic configurations, on a per formula unit (two Fe) basis,
with energies relative to the non-spin-polarized (NSP) state.}
\begin{tabular}{lcc}
\hline
 & ~~~~~$E$ (eV/f.u.)~~~~~ & ~~$N(E_F)$ (eV$^{-1}$)~~ \\
\hline
NSP &  0 &  5.45 \\
F   & -0.033 &  5.84 \\
C   &  - &  - \\
G   & - &  - \\
A   & -0.050 &  4.08 \\
A4  & -0.045 &  4.89 \\
S1  & -0.026 &  4.40 \\
S2  & -0.030 &  4.29 \\
S3  & -0.035 &  4.40 \\
X   &  - &  - \\
XZ  & -0.011 &  4.87 \\
\hline
\end{tabular}
\label{tab-mag}
\end{table}

As anticipated from the value of $N(E_F)$, which exceeds the
Stoner criterion, magnetism is expected at the DFT level.
We do find a ferromagnetic instability, as expected. The calculated
spin magnetization is 1.39 $\mu_B$ per two Fe atom unit cell.
This comes from a spin moment of 0.75 $\mu_B$ per Fe
(as measured by the magnetization in the Fe LAPW sphere, radius 2.4 bohr)
partly compensated by a back polarization on Si.
This is, however, not the calculated ground state.

We did calculations for several different possible orderings,
as summarized in Table \ref{tab-mag}.
These were ferromagnetic (F),
C-type order (C), which is a checkerboard
antiferromagnetism in the Fe plane, with like spin Fe stacked
on top of each other to make ferromagnetic chains in the $c$-axis direction,
G-type order (G), which is checkerboard in plane stacked antiferromagnetically
in the $c$ direction, and
A-type order (A), which is ferromagnetic F planes
stacked antiferromagnetically.
We also did calculations for a double period A type order (A4), consisting
of double ferromagnetic Fe layers, stacked antiferromagnetically
(...UUDDUUDD... along $c$), stripe type chain order in the Fe-planes,
as in the Fe-pnictide superconductors, stacked ferromagnetically along $c$,
(S1), stacked to run at 90$^\circ$ in alternating planes (S2),
and stacked antiferromagnetically along $c$ (S3).
Finally, we consider a double stripe order, which in plane is an $X$-point
order (ferromagnetic Fe chains running diagonally
with respect to the Fe square lattice), stacked ferromagnetically along
$c$ (X) and antiferromagnetically along $c$ (XZ). Stable solutions
were not found for G, C or X order, and instead imposing these ordering
patterns amounted to the non-spin-polarized state.

The results (Table \ref{tab-mag}) show that energy differences between
different ordering patterns and the energy difference between the
non-spin-polarized state and the ground state are comparable, and some
orders do not have solutions at all.
This is a signature of itinerant magnetism, in the sense that
there are not stable atomic
moments that exist independent of the ordering.

The lowest energy orderings are the A-type and A4-type, with the
A-type lower. The lowest energy A-type order also gives the lowest
$N(E_F)$, which might suggest a role for electrons at the Fermi energy
in stabilizing it. However, examining the other states, there is no
clear trend between $N(E_F)$ and energy among the other
antiferromagnetic orderings.

The energies show antiferromagnetic stackings along $c$ are favored,
but that this is not representable in terms of a single $c$-axis
exchange constant. For example, the energy difference between the 
A-type and ferromagnetic orders, which differ flipping every $c$-direction
bond from ferromagnetic to antiferromagnetic is $\sim$27 meV/formula unit,
while for the stripe order (S1 - S3) this
difference is only 9 meV/formula unit.
Finally, and most remarkably, while the F -- A energy difference suggests
a high energy cost for making ferromagnetic stacking, the A4 structure,
which has ferromagnetic sheets with half the c-direction bonds ferromagnetic
and half antiferromagnetic has an energy only 5 meV per formula unit above
the ground state.
The energies therefore suggest an important role for band structure and
itinerant electrons in the magnetism.

\begin{figure}
\includegraphics[width=\columnwidth]{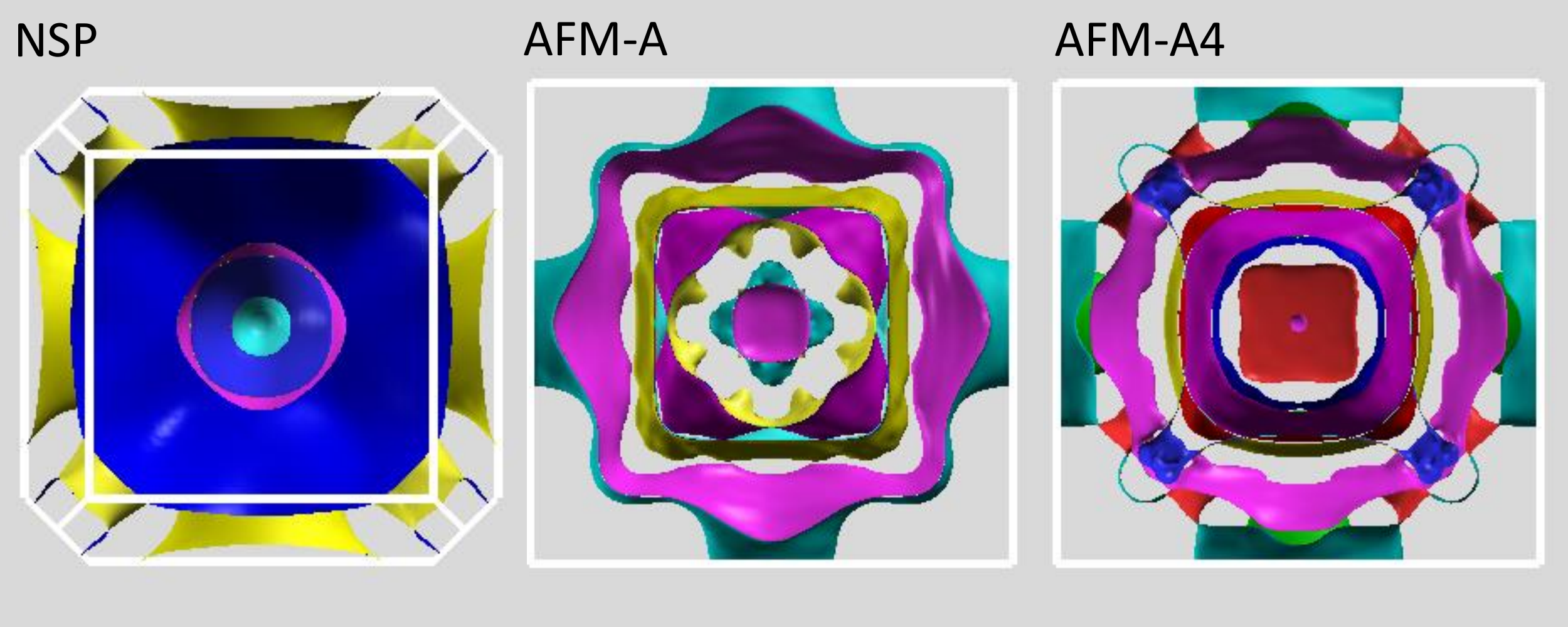}
\caption{$c$-axis view of the Fermi surface for the non-spin-polarized case,
and the A-type and A4 (see text) orders.}
\label{fermi-both}
\end{figure}

Returning to the band structure, a two fold degenerate heavy band and
a single degenerate band cross $E_F$ at almost the same point along the
$\Gamma$-$Z$ line. These three bands correspond to the h2,h3 and h4 Fermi
surfaces. The crossing is at $\sim$0.19 of the $\Gamma$-$Z$ distance.
Noting the flattened end of the h3 surface and the flattened disk shape of h4,
there is an implied nesting at a distance of $\sim$0.37$\times$$2\pi/c$.
This is intermediate between the periodicity of the A and A4 magnetic
structures (0.5 and 0.25, respectively).
Fig. \ref{fermi-both} shows a view along $k_z$ of the Fermi surfaces
for the A and A4 magnetic states in comparison with that of the 
NSP calculation. As seen, these magnetic orders gap away the large parts
of these hole sections, in particular producing new reconstructed cylindrical
sections from the large hole pancake (h4). This suggests that spin fluctuations
associated with this order would affect transport in the $c$-axis direction
more strongly than in the plane, leading to a
disproportionate reduction in conductivity along $c$ and a temperature
dependent conductivity anisotropy.

This provides an explanation for the stability of these two magnetic
structures. It also resolves an experimental puzzle regarding the
magnetic structure of the rare earth substituted compounds.
In particular, neutron diffraction experiments on NdFe$_2$Si$_2$
found Nd moments ordered with a ferromagnetic in plane order and a
$c$-axis ...UUDDUU... order, which was not readily understood in
terms of reasonable superexchange pictures. \cite{pinto}
However, Fermi surface nesting similar to what we find, along
with a slight shift in the nesting vector
closer to 0.25 could readily explain this pattern.

\begin{figure}
\includegraphics[width=\columnwidth]{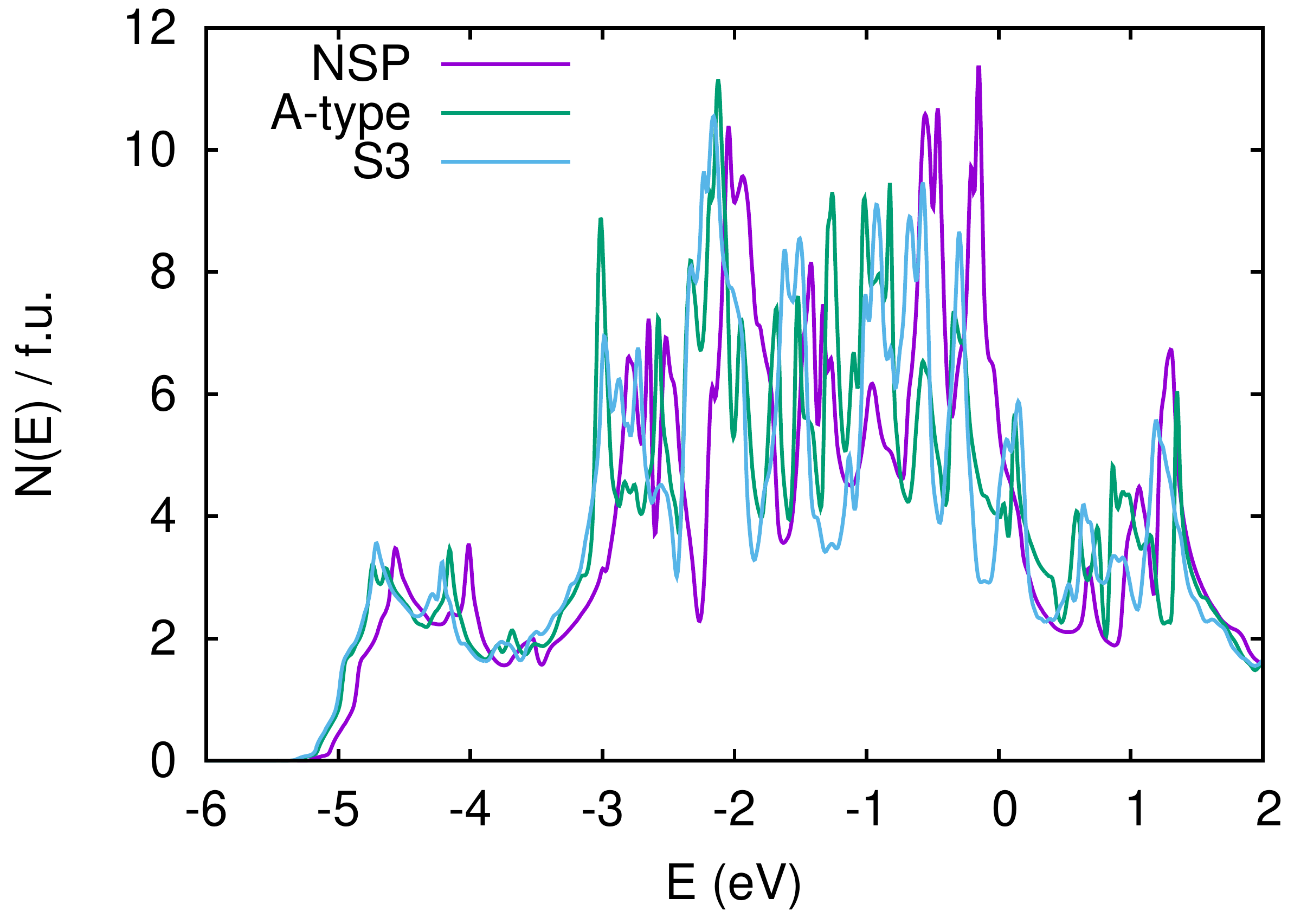}
\caption{Density of states for non-spin-polarized, the
ground state A-type antiferromagnetic state, and the S3 state.}
\label{dos-three}
\end{figure}

Fig. \ref{dos-three} shows the density of states, comparing the
NSP calculations with the lowest energy A-type antiferromagnetic
case and the stripe ordered S3 case.
As seen, the density of states is reconstructed over the energy
range of the Fe $d$ bands, i.e. -3 eV to 2 eV, on going from
the NSP state to either of the antiferromagnetic states plotted.
This is a characteristic of transition metal magnets and reflects the
coupling of the $d$ orbitals. It is the basis of the Stoner theory,
which assumes rigid shifts of the $d$ bands on forming ferromagnetic
moments. \cite{stoner,janak,andersen-es,krasko-es}
It is a feature of both local moment and itinerant transition metal
magnets, and is seen for example in the itinerant antiferromagnet Cr.
\cite{skriver}
Importantly, comparing the A-type and S3 densities of states, it is
clear that they similarly differ over the range of the $d$ bands.
This is a signature of the importance of band structure effects in the
magnetism, including both bands
near $E_F$ and bands at energies away from the Fermi level.
This is similar to the Fe-based superconductors.
\cite{johannes-fe}

Thus, at the DFT level, YFe$_2$Si$_2$ is a magnetic compound, with a
ground state having ferromagnetic layers, stacked antiferromagnetically,
and with a substantial itinerant character. The magnetic order couples
strongly to sections of the Fermi surface, particularly the largest hole
sheets. The related A4 structure is close in energy, and these compete
with a strip magnetic order analogous to that of the Fe-based
superconductors. In plane checkerboard antiferromagnetism is strongly
disfavored. Experimentally, on the other hand no magnetic order is 
reported.
This discrepancy is similar to YFe$_2$Ge$_2$,
\cite{subedi-yfe2ge2,singh-yfe2ge2,sirica}
which is an unconventional superconductor and
appears to be near a magnetic quantum critical point (note that
LuFe$_2$Ge$_2$ is magnetically ordered).
\cite{zou,fujiwara}

The situation for YFe$_2$Si$_2$ is less clear, perhaps in part because
of sample differences.
M\"ossbauer experiments for the iso-structural $R$Fe$_2$Si$_2$
compounds ($R$=rare earth), indicated the Fe has no moment in these compounds.
However, broadenings were seen in the $^{57}$Fe spectra. These were
interpreted as originating in induced Fe moments from the rare earth site.
\cite{umarji,noakes}
Detailed studies show smooth variations in lattice parameters and 
M\"ossbauer spectra across the series, similar to the
$R$Fe$_2$Ge$_2$ compounds. \cite{umarji,noakes,bara}
The $R$=La,Y,Lu compounds, where there are no rare earth moments,
are reported to be ordinary Pauli paramagnetic metals.
\cite{umarji,dommann,ijjaali}
On the other hand, magnetic behavior is readily inducted by Cr alloying,
\cite{ijjaali} and
there are reports that
show evidence for multiple Fe sites, with magnetic behavior on a portion of 
the Fe perhaps associated with Fe-Si disorder.
\cite{sankar,felner-ssc}
Other unusual magnetic behavior has also been reported under pressure and
with doping.
\cite{felner-2,felner-3}
In any case, it is clear that experiment does not show behavior consistent
with the A-type magnetic order predicted in DFT calculations.

\section{Discussion and Conclusions}

The electronic structure of metallic
YFe$_2$Si$_2$ is found to be very similar to
YFe$_2$Ge$_2$, including the qualitative structure of the Fermi surface
and the band character. Importantly, DFT calculations predict an
antiferromagnetic A-type ground state, competing with stripe order.
This is in contrast to experiment, which shows Pauli paramagnetism.
This type of discrepancy in which standard DFT calculations overestimate
the tendency towards magnetism is unusual. More commonly DFT calculations
underestimate the tendency towards moment formation as is the case
in several classes of strongly correlated materials, including
cuprates. Overestimation of the tendency to magnetism in DFT calculations
is however a characteristic of the Fe-based superconductors.
\cite{mazin-mag}
It is found as
well in metals near quantum critical points associated with itinerant 
magnetism. In that case the
discrepancy is a consequence of renormalization by
spin-fluctuations associated with the critical point and not included in
standard DFT calculations.
\cite{moriya-book,mazin-04,shimizu}

The behavior found for YFe$_2$Si$_2$ is similar to that found for
YFe$_2$Ge$_2$, except that (1) the magnetic energy scale is lower in
the silicide and (2) the ordering of the magnetic states more strongly
favors the A type ordering relative to ferromagnetism or the stripe
orders. It will be of interest to attempt synthesis of high quality well
ordered crystals of YFe$_2$Si$_2$ in order to measure its physical
properties in detail, especially specific heat and transport to assess
the extent to which magnetic fluctuations influence its properties.
In addition, spectroscopic
experiments on Fe-based superconductors and YFe$_2$Ge$_2$
have shown evidence of quantum spin fluctuations in splittings of
the Fe 3s core level under ambient conditions. \cite{bondino,sirica}
Similar experiments for YFe$_2$Si$_2$ would be of interest.

\bibliography{YFe2Si2}

\begin{thebibliography}{47}%
\makeatletter
\providecommand \@ifxundefined [1]{%
 \@ifx{#1\undefined}
}%
\providecommand \@ifnum [1]{%
 \ifnum #1\expandafter \@firstoftwo
 \else \expandafter \@secondoftwo
 \fi
}%
\providecommand \@ifx [1]{%
 \ifx #1\expandafter \@firstoftwo
 \else \expandafter \@secondoftwo
 \fi
}%
\providecommand \natexlab [1]{#1}%
\providecommand \enquote  [1]{``#1''}%
\providecommand \bibnamefont  [1]{#1}%
\providecommand \bibfnamefont [1]{#1}%
\providecommand \citenamefont [1]{#1}%
\providecommand \href@noop [0]{\@secondoftwo}%
\providecommand \href [0]{\begingroup \@sanitize@url \@href}%
\providecommand \@href[1]{\@@startlink{#1}\@@href}%
\providecommand \@@href[1]{\endgroup#1\@@endlink}%
\providecommand \@sanitize@url [0]{\catcode `\\12\catcode `\$12\catcode
  `\&12\catcode `\#12\catcode `\^12\catcode `\_12\catcode `\%12\relax}%
\providecommand \@@startlink[1]{}%
\providecommand \@@endlink[0]{}%
\providecommand \url  [0]{\begingroup\@sanitize@url \@url }%
\providecommand \@url [1]{\endgroup\@href {#1}{\urlprefix }}%
\providecommand \urlprefix  [0]{URL }%
\providecommand \Eprint [0]{\href }%
\providecommand \doibase [0]{http://dx.doi.org/}%
\providecommand \selectlanguage [0]{\@gobble}%
\providecommand \bibinfo  [0]{\@secondoftwo}%
\providecommand \bibfield  [0]{\@secondoftwo}%
\providecommand \translation [1]{[#1]}%
\providecommand \BibitemOpen [0]{}%
\providecommand \bibitemStop [0]{}%
\providecommand \bibitemNoStop [0]{.\EOS\space}%
\providecommand \EOS [0]{\spacefactor3000\relax}%
\providecommand \BibitemShut  [1]{\csname bibitem#1\endcsname}%
\let\auto@bib@innerbib\@empty
\bibitem [{\citenamefont {Johnston}(2010)}]{johnston}%
  \BibitemOpen
  \bibfield  {author} {\bibinfo {author} {\bibfnamefont {D.~C.}\ \bibnamefont
  {Johnston}},\ }\href@noop {} {\bibfield  {journal} {\bibinfo  {journal} {Adv.
  Phys.}\ }\textbf {\bibinfo {volume} {59}},\ \bibinfo {pages} {803} (\bibinfo
  {year} {2010})}\BibitemShut {NoStop}%
\bibitem [{\citenamefont {Stewart}(2011)}]{stewart}%
  \BibitemOpen
  \bibfield  {author} {\bibinfo {author} {\bibfnamefont {G.~R.}\ \bibnamefont
  {Stewart}},\ }\href@noop {} {\bibfield  {journal} {\bibinfo  {journal} {Rev.
  Mod. Phys.}\ }\textbf {\bibinfo {volume} {83}},\ \bibinfo {pages} {1589}
  (\bibinfo {year} {2011})}\BibitemShut {NoStop}%
\bibitem [{\citenamefont {Lumsden}\ and\ \citenamefont
  {Christianson}(2010)}]{lumsden}%
  \BibitemOpen
  \bibfield  {author} {\bibinfo {author} {\bibfnamefont {M.~D.}\ \bibnamefont
  {Lumsden}}\ and\ \bibinfo {author} {\bibfnamefont {A.~D.}\ \bibnamefont
  {Christianson}},\ }\href@noop {} {\bibfield  {journal} {\bibinfo  {journal}
  {J. Phys. Condens. Matter}\ }\textbf {\bibinfo {volume} {22}},\ \bibinfo
  {pages} {203203} (\bibinfo {year} {2010})}\BibitemShut {NoStop}%
\bibitem [{\citenamefont {Mazin}\ \emph
  {et~al.}(2008{\natexlab{a}})\citenamefont {Mazin}, \citenamefont {Johannes},
  \citenamefont {Boeri}, \citenamefont {Koepernik},\ and\ \citenamefont
  {Singh}}]{mazin-mag}%
  \BibitemOpen
  \bibfield  {author} {\bibinfo {author} {\bibfnamefont {I.~I.}\ \bibnamefont
  {Mazin}}, \bibinfo {author} {\bibfnamefont {M.~D.}\ \bibnamefont {Johannes}},
  \bibinfo {author} {\bibfnamefont {L.}~\bibnamefont {Boeri}}, \bibinfo
  {author} {\bibfnamefont {K.}~\bibnamefont {Koepernik}}, \ and\ \bibinfo
  {author} {\bibfnamefont {D.~J.}\ \bibnamefont {Singh}},\ }\href@noop {}
  {\bibfield  {journal} {\bibinfo  {journal} {Phys. Rev. B}\ }\textbf {\bibinfo
  {volume} {78}},\ \bibinfo {pages} {085104} (\bibinfo {year}
  {2008}{\natexlab{a}})}\BibitemShut {NoStop}%
\bibitem [{\citenamefont {Bondino}\ \emph {et~al.}(2008)\citenamefont
  {Bondino}, \citenamefont {Magnano}, \citenamefont {Malvestuto}, \citenamefont
  {Parmigiani}, \citenamefont {McGuire}, \citenamefont {Sefat}, \citenamefont
  {Sales}, \citenamefont {Jin}, \citenamefont {Mandrus}, \citenamefont
  {Plummer}, \citenamefont {Singh},\ and\ \citenamefont {Mannella}}]{bondino}%
  \BibitemOpen
  \bibfield  {author} {\bibinfo {author} {\bibfnamefont {F.}~\bibnamefont
  {Bondino}}, \bibinfo {author} {\bibfnamefont {E.}~\bibnamefont {Magnano}},
  \bibinfo {author} {\bibfnamefont {M.}~\bibnamefont {Malvestuto}}, \bibinfo
  {author} {\bibfnamefont {F.}~\bibnamefont {Parmigiani}}, \bibinfo {author}
  {\bibfnamefont {M.~A.}\ \bibnamefont {McGuire}}, \bibinfo {author}
  {\bibfnamefont {A.~S.}\ \bibnamefont {Sefat}}, \bibinfo {author}
  {\bibfnamefont {B.~C.}\ \bibnamefont {Sales}}, \bibinfo {author}
  {\bibfnamefont {R.}~\bibnamefont {Jin}}, \bibinfo {author} {\bibfnamefont
  {D.}~\bibnamefont {Mandrus}}, \bibinfo {author} {\bibfnamefont {E.~W.}\
  \bibnamefont {Plummer}}, \bibinfo {author} {\bibfnamefont {D.~J.}\
  \bibnamefont {Singh}}, \ and\ \bibinfo {author} {\bibfnamefont
  {N.}~\bibnamefont {Mannella}},\ }\href@noop {} {\bibfield  {journal}
  {\bibinfo  {journal} {Phys. Rev. Lett.}\ }\textbf {\bibinfo {volume} {101}},\
  \bibinfo {pages} {267001} (\bibinfo {year} {2008})}\BibitemShut {NoStop}%
\bibitem [{\citenamefont {Scalapino}\ \emph {et~al.}(1986)\citenamefont
  {Scalapino}, \citenamefont {Loh},\ and\ \citenamefont {Hirsch}}]{scalapino}%
  \BibitemOpen
  \bibfield  {author} {\bibinfo {author} {\bibfnamefont {D.~J.}\ \bibnamefont
  {Scalapino}}, \bibinfo {author} {\bibfnamefont {E.}~\bibnamefont {Loh}}, \
  and\ \bibinfo {author} {\bibfnamefont {J.~E.}\ \bibnamefont {Hirsch}},\
  }\href@noop {} {\bibfield  {journal} {\bibinfo  {journal} {Phys. Rev. B}\
  }\textbf {\bibinfo {volume} {34}},\ \bibinfo {pages} {8190} (\bibinfo {year}
  {1986})}\BibitemShut {NoStop}%
\bibitem [{\citenamefont {Mathur}\ \emph {et~al.}(1998)\citenamefont {Mathur},
  \citenamefont {Grosche}, \citenamefont {Julian}, \citenamefont {Walker},
  \citenamefont {Freye}, \citenamefont {Haselwimmer},\ and\ \citenamefont
  {Lonzarich}}]{mathur}%
  \BibitemOpen
  \bibfield  {author} {\bibinfo {author} {\bibfnamefont {N.~D.}\ \bibnamefont
  {Mathur}}, \bibinfo {author} {\bibfnamefont {F.~M.}\ \bibnamefont {Grosche}},
  \bibinfo {author} {\bibfnamefont {S.~R.}\ \bibnamefont {Julian}}, \bibinfo
  {author} {\bibfnamefont {I.~R.}\ \bibnamefont {Walker}}, \bibinfo {author}
  {\bibfnamefont {D.~M.}\ \bibnamefont {Freye}}, \bibinfo {author}
  {\bibfnamefont {R.~K.~W.}\ \bibnamefont {Haselwimmer}}, \ and\ \bibinfo
  {author} {\bibfnamefont {G.~G.}\ \bibnamefont {Lonzarich}},\ }\href@noop {}
  {\bibfield  {journal} {\bibinfo  {journal} {Nature}\ }\textbf {\bibinfo
  {volume} {394}},\ \bibinfo {pages} {39} (\bibinfo {year} {1998})}\BibitemShut
  {NoStop}%
\bibitem [{\citenamefont {Johannes}\ \emph {et~al.}(2004)\citenamefont
  {Johannes}, \citenamefont {Mazin}, \citenamefont {Singh},\ and\ \citenamefont
  {Papaconstantopoulos}}]{johannes}%
  \BibitemOpen
  \bibfield  {author} {\bibinfo {author} {\bibfnamefont {M.~D.}\ \bibnamefont
  {Johannes}}, \bibinfo {author} {\bibfnamefont {I.~I.}\ \bibnamefont {Mazin}},
  \bibinfo {author} {\bibfnamefont {D.~J.}\ \bibnamefont {Singh}}, \ and\
  \bibinfo {author} {\bibfnamefont {D.~A.}\ \bibnamefont
  {Papaconstantopoulos}},\ }\href@noop {} {\bibfield  {journal} {\bibinfo
  {journal} {Phys. Rev. Lett.}\ }\textbf {\bibinfo {volume} {93}},\ \bibinfo
  {pages} {097005} (\bibinfo {year} {2004})}\BibitemShut {NoStop}%
\bibitem [{\citenamefont {Moriya}(2006)}]{moriya}%
  \BibitemOpen
  \bibfield  {author} {\bibinfo {author} {\bibfnamefont {T.}~\bibnamefont
  {Moriya}},\ }\href@noop {} {\bibfield  {journal} {\bibinfo  {journal} {Proc.
  Jpn. Acad. B}\ }\textbf {\bibinfo {volume} {82}},\ \bibinfo {pages} {1}
  (\bibinfo {year} {2006})}\BibitemShut {NoStop}%
\bibitem [{\citenamefont {Mazin}\ \emph
  {et~al.}(2008{\natexlab{b}})\citenamefont {Mazin}, \citenamefont {Singh},
  \citenamefont {Johannes},\ and\ \citenamefont {Du}}]{mazin-spm}%
  \BibitemOpen
  \bibfield  {author} {\bibinfo {author} {\bibfnamefont {I.~I.}\ \bibnamefont
  {Mazin}}, \bibinfo {author} {\bibfnamefont {D.~J.}\ \bibnamefont {Singh}},
  \bibinfo {author} {\bibfnamefont {M.~D.}\ \bibnamefont {Johannes}}, \ and\
  \bibinfo {author} {\bibfnamefont {M.~H.}\ \bibnamefont {Du}},\ }\href@noop {}
  {\bibfield  {journal} {\bibinfo  {journal} {Phys. Rev. Lett.}\ }\textbf
  {\bibinfo {volume} {101}},\ \bibinfo {pages} {057003} (\bibinfo {year}
  {2008}{\natexlab{b}})}\BibitemShut {NoStop}%
\bibitem [{\citenamefont {Kuroki}\ \emph {et~al.}(2008)\citenamefont {Kuroki},
  \citenamefont {Onari}, \citenamefont {Arita}, \citenamefont {Usui},
  \citenamefont {Tanaka}, \citenamefont {Kontani},\ and\ \citenamefont
  {Aoki}}]{kuroki-spm}%
  \BibitemOpen
  \bibfield  {author} {\bibinfo {author} {\bibfnamefont {K.}~\bibnamefont
  {Kuroki}}, \bibinfo {author} {\bibfnamefont {S.}~\bibnamefont {Onari}},
  \bibinfo {author} {\bibfnamefont {R.}~\bibnamefont {Arita}}, \bibinfo
  {author} {\bibfnamefont {H.}~\bibnamefont {Usui}}, \bibinfo {author}
  {\bibfnamefont {Y.}~\bibnamefont {Tanaka}}, \bibinfo {author} {\bibfnamefont
  {H.}~\bibnamefont {Kontani}}, \ and\ \bibinfo {author} {\bibfnamefont
  {H.}~\bibnamefont {Aoki}},\ }\href@noop {} {\bibfield  {journal} {\bibinfo
  {journal} {Phys. Rev. Lett.}\ }\textbf {\bibinfo {volume} {101}},\ \bibinfo
  {pages} {087004} (\bibinfo {year} {2008})}\BibitemShut {NoStop}%
\bibitem [{\citenamefont {Scalapino}(2012)}]{scalapino2}%
  \BibitemOpen
  \bibfield  {author} {\bibinfo {author} {\bibfnamefont {D.~J.}\ \bibnamefont
  {Scalapino}},\ }\href@noop {} {\bibfield  {journal} {\bibinfo  {journal}
  {Rev. Mod. Phys.}\ }\textbf {\bibinfo {volume} {84}},\ \bibinfo {pages}
  {1383} (\bibinfo {year} {2012})}\BibitemShut {NoStop}%
\bibitem [{\citenamefont {Singh}\ and\ \citenamefont {Du}(2008)}]{singh-du}%
  \BibitemOpen
  \bibfield  {author} {\bibinfo {author} {\bibfnamefont {D.~J.}\ \bibnamefont
  {Singh}}\ and\ \bibinfo {author} {\bibfnamefont {M.~H.}\ \bibnamefont {Du}},\
  }\href@noop {} {\bibfield  {journal} {\bibinfo  {journal} {Phys. Rev. Lett.}\
  }\textbf {\bibinfo {volume} {100}},\ \bibinfo {pages} {237003} (\bibinfo
  {year} {2008})}\BibitemShut {NoStop}%
\bibitem [{\citenamefont {Guo}\ \emph {et~al.}(2010)\citenamefont {Guo},
  \citenamefont {Jin}, \citenamefont {Wang}, \citenamefont {Wang},
  \citenamefont {Zhu}, \citenamefont {Zhou}, \citenamefont {He},\ and\
  \citenamefont {Chen}}]{guo}%
  \BibitemOpen
  \bibfield  {author} {\bibinfo {author} {\bibfnamefont {J.}~\bibnamefont
  {Guo}}, \bibinfo {author} {\bibfnamefont {S.}~\bibnamefont {Jin}}, \bibinfo
  {author} {\bibfnamefont {G.}~\bibnamefont {Wang}}, \bibinfo {author}
  {\bibfnamefont {S.}~\bibnamefont {Wang}}, \bibinfo {author} {\bibfnamefont
  {K.}~\bibnamefont {Zhu}}, \bibinfo {author} {\bibfnamefont {T.}~\bibnamefont
  {Zhou}}, \bibinfo {author} {\bibfnamefont {M.}~\bibnamefont {He}}, \ and\
  \bibinfo {author} {\bibfnamefont {X.}~\bibnamefont {Chen}},\ }\href@noop {}
  {\bibfield  {journal} {\bibinfo  {journal} {Phys. Rev. B}\ }\textbf {\bibinfo
  {volume} {82}},\ \bibinfo {pages} {180520} (\bibinfo {year}
  {2010})}\BibitemShut {NoStop}%
\bibitem [{\citenamefont {Fang}\ \emph {et~al.}(2011)\citenamefont {Fang},
  \citenamefont {Wang}, \citenamefont {Dong}, \citenamefont {Li}, \citenamefont
  {Feng},\ and\ \citenamefont {Yuan}}]{fang}%
  \BibitemOpen
  \bibfield  {author} {\bibinfo {author} {\bibfnamefont {M.~H.}\ \bibnamefont
  {Fang}}, \bibinfo {author} {\bibfnamefont {H.~D.}\ \bibnamefont {Wang}},
  \bibinfo {author} {\bibfnamefont {C.~H.}\ \bibnamefont {Dong}}, \bibinfo
  {author} {\bibfnamefont {Z.~J.}\ \bibnamefont {Li}}, \bibinfo {author}
  {\bibfnamefont {C.~M.}\ \bibnamefont {Feng}}, \ and\ \bibinfo {author}
  {\bibfnamefont {H.~Q.}\ \bibnamefont {Yuan}},\ }\href@noop {} {\bibfield
  {journal} {\bibinfo  {journal} {Europhys. Lett.}\ }\textbf {\bibinfo {volume}
  {94}},\ \bibinfo {pages} {27009} (\bibinfo {year} {2011})}\BibitemShut
  {NoStop}%
\bibitem [{\citenamefont {Zhang}\ and\ \citenamefont {Singh}(2009)}]{zhang-tl}%
  \BibitemOpen
  \bibfield  {author} {\bibinfo {author} {\bibfnamefont {L.}~\bibnamefont
  {Zhang}}\ and\ \bibinfo {author} {\bibfnamefont {D.~J.}\ \bibnamefont
  {Singh}},\ }\href@noop {} {\bibfield  {journal} {\bibinfo  {journal} {Phys.
  Rev. B}\ }\textbf {\bibinfo {volume} {79}},\ \bibinfo {pages} {094528}
  (\bibinfo {year} {2009})}\BibitemShut {NoStop}%
\bibitem [{\citenamefont {Qian}\ \emph {et~al.}(2011)\citenamefont {Qian},
  \citenamefont {Wang}, \citenamefont {Jin}, \citenamefont {Zhang},
  \citenamefont {Richard}, \citenamefont {Xu}, \citenamefont {Dai},
  \citenamefont {Fang}, \citenamefont {Guo}, \citenamefont {Chen},\ and\
  \citenamefont {Ding}}]{qian}%
  \BibitemOpen
  \bibfield  {author} {\bibinfo {author} {\bibfnamefont {T.}~\bibnamefont
  {Qian}}, \bibinfo {author} {\bibfnamefont {X.~P.}\ \bibnamefont {Wang}},
  \bibinfo {author} {\bibfnamefont {W.~C.}\ \bibnamefont {Jin}}, \bibinfo
  {author} {\bibfnamefont {P.}~\bibnamefont {Zhang}}, \bibinfo {author}
  {\bibfnamefont {P.}~\bibnamefont {Richard}}, \bibinfo {author} {\bibfnamefont
  {G.}~\bibnamefont {Xu}}, \bibinfo {author} {\bibfnamefont {X.}~\bibnamefont
  {Dai}}, \bibinfo {author} {\bibfnamefont {Z.}~\bibnamefont {Fang}}, \bibinfo
  {author} {\bibfnamefont {J.~G.}\ \bibnamefont {Guo}}, \bibinfo {author}
  {\bibfnamefont {X.~L.}\ \bibnamefont {Chen}}, \ and\ \bibinfo {author}
  {\bibfnamefont {H.}~\bibnamefont {Ding}},\ }\href@noop {} {\bibfield
  {journal} {\bibinfo  {journal} {Phys. Rev. Lett.}\ }\textbf {\bibinfo
  {volume} {106}},\ \bibinfo {pages} {187001} (\bibinfo {year}
  {2011})}\BibitemShut {NoStop}%
\bibitem [{\citenamefont {Ye}\ \emph {et~al.}(2014)\citenamefont {Ye},
  \citenamefont {Zhang}, \citenamefont {Chen}, \citenamefont {Xu},
  \citenamefont {Jiang}, \citenamefont {Niu}, \citenamefont {Wen},
  \citenamefont {Xing}, \citenamefont {Wang}, \citenamefont {Jin},
  \citenamefont {Xie},\ and\ \citenamefont {Feng}}]{ye}%
  \BibitemOpen
  \bibfield  {author} {\bibinfo {author} {\bibfnamefont {Z.~R.}\ \bibnamefont
  {Ye}}, \bibinfo {author} {\bibfnamefont {Y.}~\bibnamefont {Zhang}}, \bibinfo
  {author} {\bibfnamefont {F.}~\bibnamefont {Chen}}, \bibinfo {author}
  {\bibfnamefont {M.}~\bibnamefont {Xu}}, \bibinfo {author} {\bibfnamefont
  {J.}~\bibnamefont {Jiang}}, \bibinfo {author} {\bibfnamefont {X.~H.}\
  \bibnamefont {Niu}}, \bibinfo {author} {\bibfnamefont {C.~H.~P.}\
  \bibnamefont {Wen}}, \bibinfo {author} {\bibfnamefont {L.~Y.}\ \bibnamefont
  {Xing}}, \bibinfo {author} {\bibfnamefont {X.~C.}\ \bibnamefont {Wang}},
  \bibinfo {author} {\bibfnamefont {C.~Q.}\ \bibnamefont {Jin}}, \bibinfo
  {author} {\bibfnamefont {B.~P.}\ \bibnamefont {Xie}}, \ and\ \bibinfo
  {author} {\bibfnamefont {D.~L.}\ \bibnamefont {Feng}},\ }\href@noop {}
  {\bibfield  {journal} {\bibinfo  {journal} {Phys. Rev. X}\ }\textbf {\bibinfo
  {volume} {4}},\ \bibinfo {pages} {031041} (\bibinfo {year}
  {2014})}\BibitemShut {NoStop}%
\bibitem [{\citenamefont {Zou}\ \emph {et~al.}(2014)\citenamefont {Zou},
  \citenamefont {Feng}, \citenamefont {Logg}, \citenamefont {Chen},
  \citenamefont {Lampronti},\ and\ \citenamefont {Grosche}}]{zou}%
  \BibitemOpen
  \bibfield  {author} {\bibinfo {author} {\bibfnamefont {Y.}~\bibnamefont
  {Zou}}, \bibinfo {author} {\bibfnamefont {Z.}~\bibnamefont {Feng}}, \bibinfo
  {author} {\bibfnamefont {P.~W.}\ \bibnamefont {Logg}}, \bibinfo {author}
  {\bibfnamefont {J.}~\bibnamefont {Chen}}, \bibinfo {author} {\bibfnamefont
  {G.}~\bibnamefont {Lampronti}}, \ and\ \bibinfo {author} {\bibfnamefont
  {F.~M.}\ \bibnamefont {Grosche}},\ }\href@noop {} {\bibfield  {journal}
  {\bibinfo  {journal} {Phys. Stat. Sol. Rapid Res. Lett.}\ }\textbf {\bibinfo
  {volume} {8}},\ \bibinfo {pages} {928} (\bibinfo {year} {2014})}\BibitemShut
  {NoStop}%
\bibitem [{\citenamefont {Chen}\ \emph {et~al.}(2016)\citenamefont {Chen},
  \citenamefont {Semeniuk}, \citenamefont {Feng}, \citenamefont {Reiss},
  \citenamefont {Brown}, \citenamefont {Zou}, \citenamefont {Logg},
  \citenamefont {Lampronti},\ and\ \citenamefont {Grosche}}]{chen}%
  \BibitemOpen
  \bibfield  {author} {\bibinfo {author} {\bibfnamefont {J.}~\bibnamefont
  {Chen}}, \bibinfo {author} {\bibfnamefont {K.}~\bibnamefont {Semeniuk}},
  \bibinfo {author} {\bibfnamefont {Z.}~\bibnamefont {Feng}}, \bibinfo {author}
  {\bibfnamefont {P.}~\bibnamefont {Reiss}}, \bibinfo {author} {\bibfnamefont
  {P.}~\bibnamefont {Brown}}, \bibinfo {author} {\bibfnamefont
  {Y.}~\bibnamefont {Zou}}, \bibinfo {author} {\bibfnamefont {P.~W.}\
  \bibnamefont {Logg}}, \bibinfo {author} {\bibfnamefont {G.~I.}\ \bibnamefont
  {Lampronti}}, \ and\ \bibinfo {author} {\bibfnamefont {F.~M.}\ \bibnamefont
  {Grosche}},\ }\href@noop {} {\bibfield  {journal} {\bibinfo  {journal} {Phys.
  Rev. Lett.}\ }\textbf {\bibinfo {volume} {116}},\ \bibinfo {pages} {127001}
  (\bibinfo {year} {2016})}\BibitemShut {NoStop}%
\bibitem [{\citenamefont {Subedi}(2014)}]{subedi-yfe2ge2}%
  \BibitemOpen
  \bibfield  {author} {\bibinfo {author} {\bibfnamefont {A.}~\bibnamefont
  {Subedi}},\ }\href@noop {} {\bibfield  {journal} {\bibinfo  {journal} {Phys.
  Rev. B}\ }\textbf {\bibinfo {volume} {89}},\ \bibinfo {pages} {024504}
  (\bibinfo {year} {2014})}\BibitemShut {NoStop}%
\bibitem [{\citenamefont {Singh}(2014)}]{singh-yfe2ge2}%
  \BibitemOpen
  \bibfield  {author} {\bibinfo {author} {\bibfnamefont {D.~J.}\ \bibnamefont
  {Singh}},\ }\href@noop {} {\bibfield  {journal} {\bibinfo  {journal} {Phys.
  Rev. B}\ }\textbf {\bibinfo {volume} {89}},\ \bibinfo {pages} {024505}
  (\bibinfo {year} {2014})}\BibitemShut {NoStop}%
\bibitem [{\citenamefont {Rossi}\ \emph {et~al.}(1978)\citenamefont {Rossi},
  \citenamefont {Marazza},\ and\ \citenamefont {Ferro}}]{rossi}%
  \BibitemOpen
  \bibfield  {author} {\bibinfo {author} {\bibfnamefont {D.}~\bibnamefont
  {Rossi}}, \bibinfo {author} {\bibfnamefont {R.}~\bibnamefont {Marazza}}, \
  and\ \bibinfo {author} {\bibfnamefont {R.}~\bibnamefont {Ferro}},\
  }\href@noop {} {\bibfield  {journal} {\bibinfo  {journal} {J. Less Common
  Met.}\ }\textbf {\bibinfo {volume} {58}},\ \bibinfo {pages} {203} (\bibinfo
  {year} {1978})}\BibitemShut {NoStop}%
\bibitem [{\citenamefont {Umarji}\ \emph {et~al.}(1983)\citenamefont {Umarji},
  \citenamefont {Noakes}, \citenamefont {Viccaro}, \citenamefont {Shenoy},
  \citenamefont {Aldred},\ and\ \citenamefont {Niarchos}}]{umarji}%
  \BibitemOpen
  \bibfield  {author} {\bibinfo {author} {\bibfnamefont {A.~M.}\ \bibnamefont
  {Umarji}}, \bibinfo {author} {\bibfnamefont {D.~R.}\ \bibnamefont {Noakes}},
  \bibinfo {author} {\bibfnamefont {P.~J.}\ \bibnamefont {Viccaro}}, \bibinfo
  {author} {\bibfnamefont {G.~K.}\ \bibnamefont {Shenoy}}, \bibinfo {author}
  {\bibfnamefont {A.~T.}\ \bibnamefont {Aldred}}, \ and\ \bibinfo {author}
  {\bibfnamefont {D.}~\bibnamefont {Niarchos}},\ }\href@noop {} {\bibfield
  {journal} {\bibinfo  {journal} {J. Magn. Magn. Mater.}\ }\textbf {\bibinfo
  {volume} {36}},\ \bibinfo {pages} {61} (\bibinfo {year} {1983})}\BibitemShut
  {NoStop}%
\bibitem [{\citenamefont {Noakes}\ \emph {et~al.}(1983)\citenamefont {Noakes},
  \citenamefont {Umarji},\ and\ \citenamefont {Shenoy}}]{noakes}%
  \BibitemOpen
  \bibfield  {author} {\bibinfo {author} {\bibfnamefont {D.~R.}\ \bibnamefont
  {Noakes}}, \bibinfo {author} {\bibfnamefont {A.~M.}\ \bibnamefont {Umarji}},
  \ and\ \bibinfo {author} {\bibfnamefont {G.~K.}\ \bibnamefont {Shenoy}},\
  }\href@noop {} {\bibfield  {journal} {\bibinfo  {journal} {J. Magn. Magn.
  Mater.}\ }\textbf {\bibinfo {volume} {39}},\ \bibinfo {pages} {309} (\bibinfo
  {year} {1983})}\BibitemShut {NoStop}%
\bibitem [{\citenamefont {Bara}\ \emph {et~al.}(1990)\citenamefont {Bara},
  \citenamefont {Hrynkiewicz}, \citenamefont {Milos},\ and\ \citenamefont
  {Szytula}}]{bara}%
  \BibitemOpen
  \bibfield  {author} {\bibinfo {author} {\bibfnamefont {J.~J.}\ \bibnamefont
  {Bara}}, \bibinfo {author} {\bibfnamefont {H.~U.}\ \bibnamefont
  {Hrynkiewicz}}, \bibinfo {author} {\bibfnamefont {A.}~\bibnamefont {Milos}},
  \ and\ \bibinfo {author} {\bibfnamefont {A.}~\bibnamefont {Szytula}},\
  }\href@noop {} {\bibfield  {journal} {\bibinfo  {journal} {J. Less Common
  Metals}\ }\textbf {\bibinfo {volume} {161}},\ \bibinfo {pages} {185}
  (\bibinfo {year} {1990})}\BibitemShut {NoStop}%
\bibitem [{\citenamefont {Dommann}\ \emph {et~al.}(1988)\citenamefont
  {Dommann}, \citenamefont {Hulliger},\ and\ \citenamefont
  {Baerlocher}}]{dommann}%
  \BibitemOpen
  \bibfield  {author} {\bibinfo {author} {\bibfnamefont {A.}~\bibnamefont
  {Dommann}}, \bibinfo {author} {\bibfnamefont {F.}~\bibnamefont {Hulliger}}, \
  and\ \bibinfo {author} {\bibfnamefont {C.}~\bibnamefont {Baerlocher}},\
  }\href@noop {} {\bibfield  {journal} {\bibinfo  {journal} {J. Less Common
  Metals}\ }\textbf {\bibinfo {volume} {138}},\ \bibinfo {pages} {113}
  (\bibinfo {year} {1988})}\BibitemShut {NoStop}%
\bibitem [{\citenamefont {Ijjaali}\ \emph {et~al.}(1998)\citenamefont
  {Ijjaali}, \citenamefont {Venturini},\ and\ \citenamefont
  {Malaman}}]{ijjaali}%
  \BibitemOpen
  \bibfield  {author} {\bibinfo {author} {\bibfnamefont {I.}~\bibnamefont
  {Ijjaali}}, \bibinfo {author} {\bibfnamefont {G.}~\bibnamefont {Venturini}},
  \ and\ \bibinfo {author} {\bibfnamefont {B.}~\bibnamefont {Malaman}},\
  }\href@noop {} {\bibfield  {journal} {\bibinfo  {journal} {J. Alloys
  Compds.}\ }\textbf {\bibinfo {volume} {279}},\ \bibinfo {pages} {102}
  (\bibinfo {year} {1998})}\BibitemShut {NoStop}%
\bibitem [{\citenamefont {Sankar}\ \emph {et~al.}(1976)\citenamefont {Sankar},
  \citenamefont {Malik}, \citenamefont {Rao},\ and\ \citenamefont
  {Obermeyer}}]{sankar}%
  \BibitemOpen
  \bibfield  {author} {\bibinfo {author} {\bibfnamefont {S.~G.}\ \bibnamefont
  {Sankar}}, \bibinfo {author} {\bibfnamefont {S.~K.}\ \bibnamefont {Malik}},
  \bibinfo {author} {\bibfnamefont {V.~U.~S.}\ \bibnamefont {Rao}}, \ and\
  \bibinfo {author} {\bibfnamefont {R.}~\bibnamefont {Obermeyer}},\ }\href@noop
  {} {\bibfield  {journal} {\bibinfo  {journal} {API Conf. Proc.}\ }\textbf
  {\bibinfo {volume} {34}},\ \bibinfo {pages} {236} (\bibinfo {year}
  {1976})}\BibitemShut {NoStop}%
\bibitem [{\citenamefont {Felner}\ \emph {et~al.}(1975)\citenamefont {Felner},
  \citenamefont {Mayer}, \citenamefont {Grill},\ and\ \citenamefont
  {Schieber}}]{felner-ssc}%
  \BibitemOpen
  \bibfield  {author} {\bibinfo {author} {\bibfnamefont {I.}~\bibnamefont
  {Felner}}, \bibinfo {author} {\bibfnamefont {I.}~\bibnamefont {Mayer}},
  \bibinfo {author} {\bibfnamefont {A.}~\bibnamefont {Grill}}, \ and\ \bibinfo
  {author} {\bibfnamefont {M.}~\bibnamefont {Schieber}},\ }\href@noop {}
  {\bibfield  {journal} {\bibinfo  {journal} {Solid State Commun.}\ }\textbf
  {\bibinfo {volume} {16}},\ \bibinfo {pages} {1005} (\bibinfo {year}
  {1975})}\BibitemShut {NoStop}%
\bibitem [{\citenamefont {Felner}\ \emph {et~al.}(2015)\citenamefont {Felner},
  \citenamefont {Lv}, \citenamefont {Zhao},\ and\ \citenamefont
  {Chu}}]{felner-2}%
  \BibitemOpen
  \bibfield  {author} {\bibinfo {author} {\bibfnamefont {I.}~\bibnamefont
  {Felner}}, \bibinfo {author} {\bibfnamefont {B.}~\bibnamefont {Lv}}, \bibinfo
  {author} {\bibfnamefont {K.}~\bibnamefont {Zhao}}, \ and\ \bibinfo {author}
  {\bibfnamefont {C.~W.}\ \bibnamefont {Chu}},\ }\href@noop {} {\bibfield
  {journal} {\bibinfo  {journal} {J. Supercond. Nov. Magn.}\ }\textbf {\bibinfo
  {volume} {28}},\ \bibinfo {pages} {1207} (\bibinfo {year}
  {2015})}\BibitemShut {NoStop}%
\bibitem [{\citenamefont {Felner}\ \emph {et~al.}(2014)\citenamefont {Felner},
  \citenamefont {Lv},\ and\ \citenamefont {Chu}}]{felner-3}%
  \BibitemOpen
  \bibfield  {author} {\bibinfo {author} {\bibfnamefont {I.}~\bibnamefont
  {Felner}}, \bibinfo {author} {\bibfnamefont {B.}~\bibnamefont {Lv}}, \ and\
  \bibinfo {author} {\bibfnamefont {C.~W.}\ \bibnamefont {Chu}},\ }\href@noop
  {} {\bibfield  {journal} {\bibinfo  {journal} {J. Phys. Condens. Matter}\
  }\textbf {\bibinfo {volume} {25}},\ \bibinfo {pages} {476002} (\bibinfo
  {year} {2014})}\BibitemShut {NoStop}%
\bibitem [{\citenamefont {Perdew}\ \emph {et~al.}(1996)\citenamefont {Perdew},
  \citenamefont {Burke},\ and\ \citenamefont {Ernzerhof}}]{pbe}%
  \BibitemOpen
  \bibfield  {author} {\bibinfo {author} {\bibfnamefont {J.~P.}\ \bibnamefont
  {Perdew}}, \bibinfo {author} {\bibfnamefont {K.}~\bibnamefont {Burke}}, \
  and\ \bibinfo {author} {\bibfnamefont {M.}~\bibnamefont {Ernzerhof}},\
  }\href@noop {} {\bibfield  {journal} {\bibinfo  {journal} {Phys. Rev. Lett.}\
  }\textbf {\bibinfo {volume} {77}},\ \bibinfo {pages} {3865} (\bibinfo {year}
  {1996})}\BibitemShut {NoStop}%
\bibitem [{\citenamefont {Singh}\ and\ \citenamefont
  {Nordstrom}(2006)}]{singh-book}%
  \BibitemOpen
  \bibfield  {author} {\bibinfo {author} {\bibfnamefont {D.~J.}\ \bibnamefont
  {Singh}}\ and\ \bibinfo {author} {\bibfnamefont {L.}~\bibnamefont
  {Nordstrom}},\ }\href@noop {} {\emph {\bibinfo {title} {{Planewaves
  Pseudopotentials and the LAPW Method, 2nd Edition}}}}\ (\bibinfo  {publisher}
  {Springer, Berlin},\ \bibinfo {year} {2006})\BibitemShut {NoStop}%
\bibitem [{\citenamefont {Blaha}\ \emph {et~al.}(2001)\citenamefont {Blaha},
  \citenamefont {Schwarz}, \citenamefont {Madsen}, \citenamefont {Kvasnicka},\
  and\ \citenamefont {Luitz}}]{wien2k}%
  \BibitemOpen
  \bibfield  {author} {\bibinfo {author} {\bibfnamefont {P.}~\bibnamefont
  {Blaha}}, \bibinfo {author} {\bibfnamefont {K.}~\bibnamefont {Schwarz}},
  \bibinfo {author} {\bibfnamefont {G.}~\bibnamefont {Madsen}}, \bibinfo
  {author} {\bibfnamefont {D.}~\bibnamefont {Kvasnicka}}, \ and\ \bibinfo
  {author} {\bibfnamefont {J.}~\bibnamefont {Luitz}},\ }\href@noop {} {\emph
  {\bibinfo {title} {WIEN2k, An Augmented Plane Wave + Local Orbitals Program
  for Calculating Crystal Properties}}}\ (\bibinfo  {publisher} {K. Schwarz,
  Tech. Univ. Wien, Austria},\ \bibinfo {year} {2001})\BibitemShut {NoStop}%
\bibitem [{\citenamefont {Stoner}(1939)}]{stoner}%
  \BibitemOpen
  \bibfield  {author} {\bibinfo {author} {\bibfnamefont {E.~C.}\ \bibnamefont
  {Stoner}},\ }\href@noop {} {\bibfield  {journal} {\bibinfo  {journal} {Proc.
  R. Soc. London Ser. A}\ }\textbf {\bibinfo {volume} {169}},\ \bibinfo {pages}
  {339} (\bibinfo {year} {1939})}\BibitemShut {NoStop}%
\bibitem [{\citenamefont {Janak}(1977)}]{janak}%
  \BibitemOpen
  \bibfield  {author} {\bibinfo {author} {\bibfnamefont {J.~F.}\ \bibnamefont
  {Janak}},\ }\href@noop {} {\bibfield  {journal} {\bibinfo  {journal} {Phys.
  Rev. B}\ }\textbf {\bibinfo {volume} {16}},\ \bibinfo {pages} {255} (\bibinfo
  {year} {1977})}\BibitemShut {NoStop}%
\bibitem [{\citenamefont {Pinto}\ and\ \citenamefont {Shaked}(1973)}]{pinto}%
  \BibitemOpen
  \bibfield  {author} {\bibinfo {author} {\bibfnamefont {H.}~\bibnamefont
  {Pinto}}\ and\ \bibinfo {author} {\bibfnamefont {H.}~\bibnamefont {Shaked}},\
  }\href@noop {} {\bibfield  {journal} {\bibinfo  {journal} {Phys. Rev. B}\
  }\textbf {\bibinfo {volume} {7}},\ \bibinfo {pages} {3261} (\bibinfo {year}
  {1973})}\BibitemShut {NoStop}%
\bibitem [{\citenamefont {Andersen}\ \emph {et~al.}(1977)\citenamefont
  {Andersen}, \citenamefont {Madsen}, \citenamefont {Poulsen}, \citenamefont
  {Jepsen},\ and\ \citenamefont {Kollar}}]{andersen-es}%
  \BibitemOpen
  \bibfield  {author} {\bibinfo {author} {\bibfnamefont {O.~K.}\ \bibnamefont
  {Andersen}}, \bibinfo {author} {\bibfnamefont {J.}~\bibnamefont {Madsen}},
  \bibinfo {author} {\bibfnamefont {U.~K.}\ \bibnamefont {Poulsen}}, \bibinfo
  {author} {\bibfnamefont {O.}~\bibnamefont {Jepsen}}, \ and\ \bibinfo {author}
  {\bibfnamefont {J.}~\bibnamefont {Kollar}},\ }\href@noop {} {\bibfield
  {journal} {\bibinfo  {journal} {Physica B}\ }\textbf {\bibinfo {volume}
  {86-88}},\ \bibinfo {pages} {249} (\bibinfo {year} {1977})}\BibitemShut
  {NoStop}%
\bibitem [{\citenamefont {Krasko}(1987)}]{krasko-es}%
  \BibitemOpen
  \bibfield  {author} {\bibinfo {author} {\bibfnamefont {G.~L.}\ \bibnamefont
  {Krasko}},\ }\href@noop {} {\bibfield  {journal} {\bibinfo  {journal} {Phys.
  Rev. B}\ }\textbf {\bibinfo {volume} {36}},\ \bibinfo {pages} {8565}
  (\bibinfo {year} {1987})}\BibitemShut {NoStop}%
\bibitem [{\citenamefont {Skriver}(1981)}]{skriver}%
  \BibitemOpen
  \bibfield  {author} {\bibinfo {author} {\bibfnamefont {H.~L.}\ \bibnamefont
  {Skriver}},\ }\href@noop {} {\bibfield  {journal} {\bibinfo  {journal} {J.
  Phys. F}\ }\textbf {\bibinfo {volume} {11}},\ \bibinfo {pages} {97} (\bibinfo
  {year} {1981})}\BibitemShut {NoStop}%
\bibitem [{\citenamefont {Johannes}\ and\ \citenamefont
  {Mazin}(2009)}]{johannes-fe}%
  \BibitemOpen
  \bibfield  {author} {\bibinfo {author} {\bibfnamefont {M.~D.}\ \bibnamefont
  {Johannes}}\ and\ \bibinfo {author} {\bibfnamefont {I.~I.}\ \bibnamefont
  {Mazin}},\ }\href@noop {} {\bibfield  {journal} {\bibinfo  {journal} {Phys.
  Rev. B}\ }\textbf {\bibinfo {volume} {79}},\ \bibinfo {pages} {220510}
  (\bibinfo {year} {2009})}\BibitemShut {NoStop}%
\bibitem [{\citenamefont {Sirica}\ \emph {et~al.}(2015)\citenamefont {Sirica},
  \citenamefont {Bondino}, \citenamefont {Nappini}, \citenamefont {Pis},
  \citenamefont {Poudel}, \citenamefont {Christianson}, \citenamefont
  {Mandrus}, \citenamefont {Singh},\ and\ \citenamefont {Mannella}}]{sirica}%
  \BibitemOpen
  \bibfield  {author} {\bibinfo {author} {\bibfnamefont {N.}~\bibnamefont
  {Sirica}}, \bibinfo {author} {\bibfnamefont {F.}~\bibnamefont {Bondino}},
  \bibinfo {author} {\bibfnamefont {S.}~\bibnamefont {Nappini}}, \bibinfo
  {author} {\bibfnamefont {I.}~\bibnamefont {Pis}}, \bibinfo {author}
  {\bibfnamefont {L.}~\bibnamefont {Poudel}}, \bibinfo {author} {\bibfnamefont
  {A.~D.}\ \bibnamefont {Christianson}}, \bibinfo {author} {\bibfnamefont
  {D.}~\bibnamefont {Mandrus}}, \bibinfo {author} {\bibfnamefont {D.~J.}\
  \bibnamefont {Singh}}, \ and\ \bibinfo {author} {\bibfnamefont
  {N.}~\bibnamefont {Mannella}},\ }\href@noop {} {\bibfield  {journal}
  {\bibinfo  {journal} {Phys. Rev. B}\ }\textbf {\bibinfo {volume} {91}},\
  \bibinfo {pages} {121102} (\bibinfo {year} {2015})}\BibitemShut {NoStop}%
\bibitem [{\citenamefont {Fujiwara}\ \emph {et~al.}(2007)\citenamefont
  {Fujiwara}, \citenamefont {Aso}, \citenamefont {Yamamoto}, \citenamefont
  {Hedo}, \citenamefont {Saiga}, \citenamefont {Nishi}, \citenamefont
  {Uwatoko},\ and\ \citenamefont {Hirota}}]{fujiwara}%
  \BibitemOpen
  \bibfield  {author} {\bibinfo {author} {\bibfnamefont {T.}~\bibnamefont
  {Fujiwara}}, \bibinfo {author} {\bibfnamefont {N.}~\bibnamefont {Aso}},
  \bibinfo {author} {\bibfnamefont {H.}~\bibnamefont {Yamamoto}}, \bibinfo
  {author} {\bibfnamefont {M.}~\bibnamefont {Hedo}}, \bibinfo {author}
  {\bibfnamefont {Y.}~\bibnamefont {Saiga}}, \bibinfo {author} {\bibfnamefont
  {M.}~\bibnamefont {Nishi}}, \bibinfo {author} {\bibfnamefont
  {Y.}~\bibnamefont {Uwatoko}}, \ and\ \bibinfo {author} {\bibfnamefont
  {K.}~\bibnamefont {Hirota}},\ }\href@noop {} {\bibfield  {journal} {\bibinfo
  {journal} {J. Phys. Soc. Jpn.}\ }\textbf {\bibinfo {volume} {76}},\ \bibinfo
  {pages} {SA60} (\bibinfo {year} {2007})}\BibitemShut {NoStop}%
\bibitem [{\citenamefont {Moriya}(1985)}]{moriya-book}%
  \BibitemOpen
  \bibfield  {author} {\bibinfo {author} {\bibfnamefont {T.}~\bibnamefont
  {Moriya}},\ }\href@noop {} {\emph {\bibinfo {title} {{Spin Fluctuations in
  Itinerant Electron Magnetism}}}}\ (\bibinfo  {publisher} {Springer, Berlin},\
  \bibinfo {year} {1985})\BibitemShut {NoStop}%
\bibitem [{\citenamefont {Mazin}\ and\ \citenamefont {Singh}(2004)}]{mazin-04}%
  \BibitemOpen
  \bibfield  {author} {\bibinfo {author} {\bibfnamefont {I.~I.}\ \bibnamefont
  {Mazin}}\ and\ \bibinfo {author} {\bibfnamefont {D.~J.}\ \bibnamefont
  {Singh}},\ }\href@noop {} {\bibfield  {journal} {\bibinfo  {journal} {Phys.
  Rev. B}\ }\textbf {\bibinfo {volume} {69}},\ \bibinfo {pages} {020402}
  (\bibinfo {year} {2004})}\BibitemShut {NoStop}%
\bibitem [{\citenamefont {Shimizu}(1981)}]{shimizu}%
  \BibitemOpen
  \bibfield  {author} {\bibinfo {author} {\bibfnamefont {M.}~\bibnamefont
  {Shimizu}},\ }\href@noop {} {\bibfield  {journal} {\bibinfo  {journal} {Rep.
  Prog. Phys.}\ }\textbf {\bibinfo {volume} {44}},\ \bibinfo {pages} {329}
  (\bibinfo {year} {1981})}\BibitemShut {NoStop}%
\end{thebibliography}%

\end{document}